\documentclass[
    ,final            % use final for the camera ready runs
  ]
  {aipproc}

\layoutstyle{6x9}

\newcommand{\MET}{\slash\!\!\!\!\!E_T}
\newcommand{\DZero}{D0$\!\!\!\!\!\slash$}

\begin{document}

\title{Higgs Boson Decay into a Pair of Leptons}


\author{Edmond L.\ Berger}{
address={High Energy Physics Division, Argonne National Laboratory,
Argonne, Illinois 60439}
}
\author{Zack Sullivan}{
address={High Energy Physics Division, Argonne National Laboratory,
Argonne, Illinois 60439}
}

\begin{abstract}
The decay of a Higgs boson into a pair of $W$ bosons 
$h \rightarrow W^+W^-$, is a
dominant mode for Higgs boson masses above 135~GeV.  At hadron colliders,
searches for this decay focus on channels in which both $W$ bosons decay
leptonically, $h \rightarrow W^+ W^- \rightarrow l^+ l^-$ plus missing energy.
We show that semileptonic decays of heavy flavors are an important background
to this signal.  Lepton isolation provides too little suppression of heavy
flavor contributions, and an additional 4 to 8 orders-of-magnitude suppression
must come from physics cuts.  An increase of the cut on the the minimum
transverse momentum of non-leading leptons in multilepton events is one
effective way to achieve the needed suppression, without appreciable loss of
the Higgs boson signal.
\end{abstract}

\maketitle

\section{Introduction}

%\indent\indent
In hadron collisions, the cleanest signature for the Higgs boson decay $h\to
W^+W^-$ is two isolated opposite-sign leptons plus missing transverse energy
($\MET$) from the neutrinos in $W\to l\nu$.  One class of reducible
backgrounds involves processes with heavy-flavor (HF) hadrons in the final
state, such as $W c$, $W b \bar{b}$, and $b \bar{b}$, where at least one
lepton comes from the decay of a HF hadron (a hadron that includes either a
bottom or charm quark).  In this report I summarize a recent
study~\cite{Sullivan:2006hb} in which we demonstrate that isolation does not
sufficiently suppress these HF backgrounds.  Rather, the size of the
heavy-flavor background is determined by details of the applied physics cuts,
and current analyses must be improved to remove this background to $h\to
W^+W^-$.

We provide a full simulation of the backgrounds for $h \to W^+W^-\to
l^+l^-{\MET}$, following the analysis chains of two studies: one by the
\DZero\ Collaboration~\cite{Abazov:2005un} from which a limit is set on $h\to
WW$ at the Tevatron, and one by the ATLAS Collaboration~\cite{ATLTDR} that
estimates the reach at the LHC. The heavy-flavor background could be
overwhelming with the default cuts at the LHC, but we propose a new, more
restrictive cut that would significantly reduce the background.  We emphasize
the value of direct measurements of the magnitude and kinematic variation the
heavy-flavor background in the Tevatron and LHC data.  Space restrictions
limit this brief summary to our LHC investigation.

\section{Heavy flavor background at the LHC}

%\indent\indent
The issue for the heavy flavor backgrounds is the extent to which lepton
isolation and subsequent kinematic physics cuts can suppress them.  The nature
of the challenge at the LHC is illustrated by a comparison of $\sigma\times
B(h \rightarrow W W^* \rightarrow l^+l^-\nu \bar{\nu}) \sim 0.7$~pb for $m_h =
150$ to $190$~GeV, with the size of $\sigma^{\rm inclusive}_{b \bar{b}} \sim 5
\times 10^8$~pb.  Isolation in $b \rightarrow l X$ even at the $0.5\%$ level
leaves a HF $l^+ l^- \MET$ background that is $10^4$ greater than the signal.
We must address questions of both the magnitude and the shape of the
backgrounds.
  
In order to make statements regarding experimental issues, we require a
detailed simulation of reconstructed events.  We run events through the PYTHIA
6.322~\cite{PYTHIA} showering Monte Carlo, and feed the output through a
heavily modified version of PGS~\cite{Carena:2000yx} that reproduces the
results of the relevant full ATLAS detector simulations to within
10\%~\cite{ATLLEPS}.  This code has a more accurate treatment of geometric
effects and efficiencies than ATLFAST. The ATLAS physics cuts described in the
ATLAS Technical Design Report (TDR)~\cite{ATLTDR} are then applied to the
objects found in PGS.  We use MadEvent 3.0~\cite{Maltoni:2002qb} to generate
hard events, and we match the cross sections after showering to the
differential next-to-leading order (NLO) cross sections.  For $Wjj$ and the
relevant single-top-quark process, a $K$ factor times a leading-order (LO)
distribution is sufficient to retain all angular correlations.  Continuum
$W^+W^-$ and $h\to W^+W^-$ are evaluated using PYTHIA routines with $K$
factors.

\begin{table}[bth]
\begin{tabular}{lccccccc}
\hline
\multicolumn{1}{c}{Cut level} &
$h\to WW$ & $WW$ &
$b\bar bj^\star$ & $Wc$ &
Single-top & $Wb\bar b$ &
$Wc\bar c$ \\ \hline
Isolated $l^+l^-$ & 336 & 1270 & $>35700$ & 12200 & 3010 & 1500 & 1110 \\
$E_{Tl_1}>20$ GeV & 324 & 1210 & $>5650$ & 11300 & 2550 & 1270 & 963 \\
$\MET > 40$ GeV & 244 & 661 & $>3280$ & 2710 & 726 & 364 & 468 \\
$M_{ll} < 80$ GeV & 240 & 376 & $>3270$ & 2450 & 692 & 320 & 461 \\
$\Delta\phi < 1.0$ & 136 & 124 & $>1670$ & 609 & 115 & 94 & 131 \\
$|\theta_{ll}|<0.9$ & 81 & 83 & $>1290$ & 393 & 68 & 49 & 115 \\
$|\eta_{l_1}-\eta_{l_2}| < 1.5$ & 76 & 71 & $>678$ & 320 & 48 & 24 & 104 \\
Jet veto & 41 & 43 & $>557$ & 175 & 11 & 12 & 7.4 \\
$130 < M_T^{ll} < 160$ GeV & 18 & 11 & {---} & 0.21 & 1.3 &
0.04 & 0.09\\\hline
\end{tabular}
\caption{Cross sections (in fb) for opposite-sign leptons as a function of
cuts for the 160 GeV Higgs boson ATLAS analysis.  $b\bar bj^{\star}$
production is a lower limit based on limited phase space.  A dash indicates
statistics are too small to estimate.}
\label{tab:atlcutline}
\end{table}
In Table \ref{tab:atlcutline} we show the cross sections we compute for 
opposite-sign dileptons. The 
second column shows the signal process $h\rightarrow W W$, 
and the remaining columns display the backgrounds that we examine, beginning 
with continuum $WW$ production and followed by several heavy flavor 
backgrounds.
The first level of cuts requires two isolated leptons, each with $p_{Tl}>10$
GeV and $|\eta_l|<2.5$.  Isolation of electrons and muons replicates recent
ATLAS descriptions~\cite{ATLTDR,ATLLEPS}, and it is applied within the
modified PGS detector simulation.  Next, a cut is placed on the transverse
energy of the reconstructed highest-$E_T$ lepton $l_1$ of $E_{Tl_1}>20$ GeV.
A fairly high missing energy of 40~GeV is then required.  Spin correlations in
$h \rightarrow W^+ W^- \rightarrow l^+ l^- \MET$ tend to send the leptons in
the same direction.  Consequently, their invariant mass is low, and ATLAS
requires $M_{ll}< 80$~GeV.  Likewise, the angle between the leptons should
also be small, and an aggressive cut is made on the azimuthal angle between
the leptons, $\Delta\phi<1.0$.  The next-to-last cut is a veto of any event
having a jet with $E_{Tj}>15$~GeV, and $|\eta_j|<3.2$.  It serves to reject
background from $t \bar{t}$ production.  Finally, a tight cut is made on the
transverse mass $M_T^{ll}$ of the dilepton and missing energy.  It appears
naively to remove most of the heavy-flavor background.

\begin{figure}[tbh]
\includegraphics[%
  width=10cm,
  height=6.7cm,
  keepaspectratio]{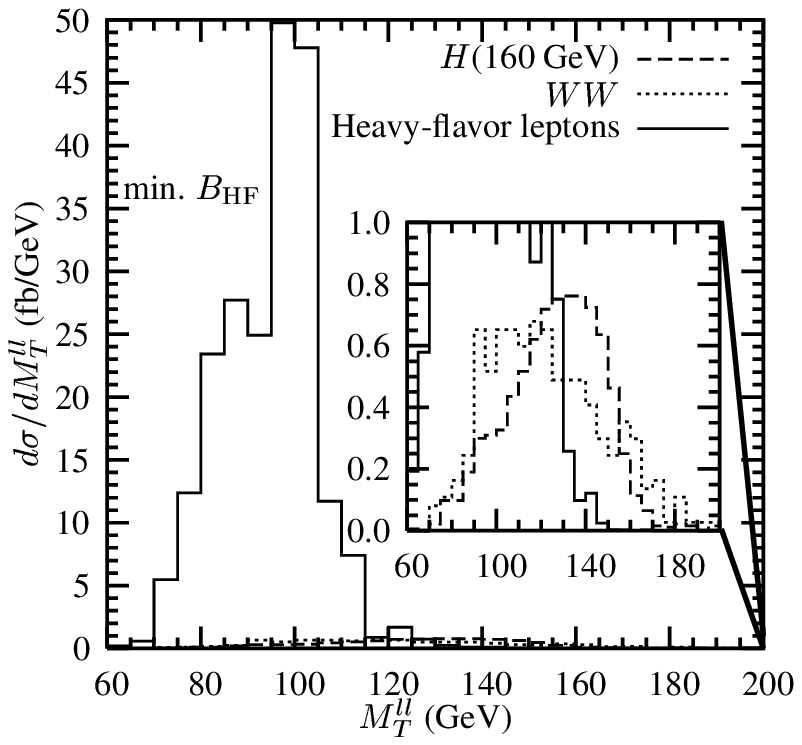}~~\includegraphics[%
  width=10cm,
  height=6.7cm,
  keepaspectratio]{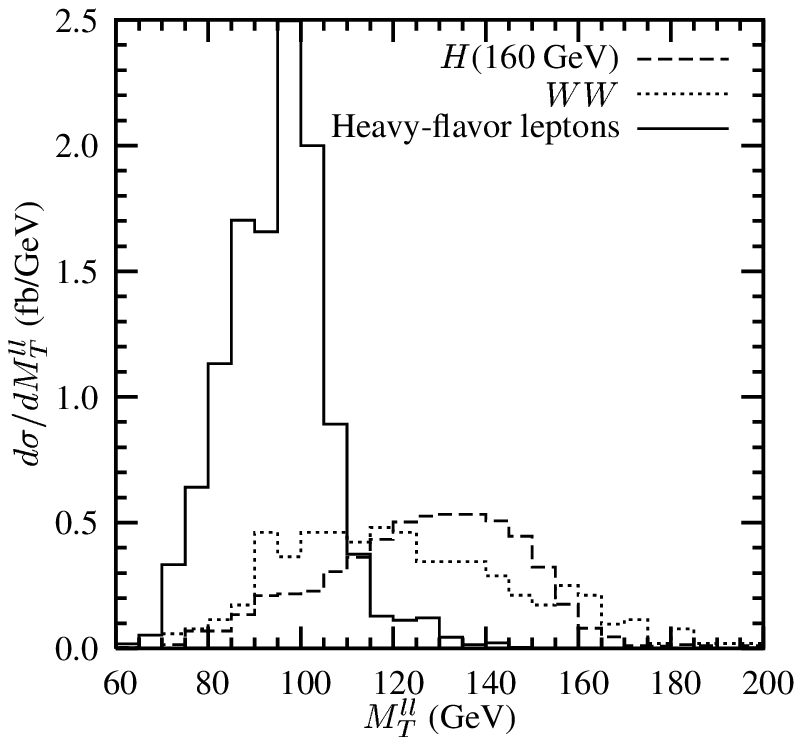}
\caption{(a) Opposite-sign dilepton transverse mass distribution for a 160~GeV
Higgs boson, the continuum $WW$ background, and the sum of heavy-flavor
backgrounds (HFB) at ATLAS.  The inset shows a blow-up of the signal region.
(b) Same as (a) but after the cut on the next-to-leading lepton $p_{Tl_2}$ is
raised from 10 GeV to 20~GeV.
\label{fig:atlmtll}}
\end{figure}

The final cut on $M_T^{ll}$ is the key to the ATLAS sensitivity.  In Fig.\
\ref{fig:atlmtll}(a) we see a comparison of the Higgs boson signal, the
continuum $WW$ background, and the heavy-flavor backgrounds.  The HF
backgrounds are more than an order of magnitude larger than the previously
calculated backgrounds for $M_T^{ll} < 110$~GeV.  As a result of the physics
cuts and lepton isolation, the $b\bar b$ background has been suppressed by 11
orders of magnitude.  It is unlikely that the tail of this distribution cuts
off sharply at 125~GeV.  It would be difficult to believe an excess observed
in the region $M_T^{ll}<160$~GeV without a measurement of this HF background.
Even for a 200~GeV Higgs boson, the median transverse mass is below 140~GeV,
leading to poor mass resolution if events are observed.

Fortunately, we can reduce the HF background to a manageable level by pushing
the $M_T^{ll}$ mass peak associated with the heavy-flavors below 110--120~GeV.
The distribution in $E_T$ of the leading lepton in the $W+$jets and single-top
samples is fairly insensitive to small increases in the $E_T$ threshold near
20~GeV.  This feature is not surprising since the leptons come from real 
$W$ decays.
The leading lepton from $b$ or $c$ decay falls faster, but an increase in the
cut will not improve the overall significance. On the other hand, the
next-to-leading lepton has an exponentially falling background as a function
of $E_T$.  An increase of the minimum transverse energy cut on additional
leptons from 10 GeV to 20~GeV reduces the background by roughly a factor of
20, while maintaining about $2/3$ of the signal and continuum $WW$
backgrounds.  In particular, the dangerous $b\bar b$ background drops by a
factor of 30, the $Wj+X$ backgrounds go down a factor of 10, and
single-top-quark production goes down a factor of 5.  Such a cut is nearly a
``magic bullet'' for Higgs boson masses above 140~GeV.  An estimate of the
effect of this one change in the cuts is shown for the signal and total
backgrounds in Fig.\ \ref{fig:atlmtll}(b).  The leading edge of the
heavy-flavor transverse-mass peak is 20~GeV lower than with the default cuts.
The downward shift of this leading edge, along with the lower overall
magnitude of the background, protects the Higgs boson signal region from
uncertainties in the modeling of the heavy-flavor background.  The residual HF
background will still be measurable at lower $M_T^{ll}$, and it provides a
control sample.

Our analysis demonstrates that despite small efficiencies, heavy-flavor decays
into leptons are a potentially serious background.  An increase in the
transverse-energy cut of secondary leptons is effective at reducing the
background, but every level of cuts is significant.  Some of the proposed cuts
are sensitive to actual detector performance, noise, and the underlying event
--- none of which will be known until data are accumulated.  Extrapolations of
the magnitude and shape of the HF background using Monte Carlo techniques have
large inherent uncertainties. We recommend that the HF background be measured
with cuts as close as possible to the final sample.  At the LHC the HF
background is large enough that it can be studied in the $M_T^{ll}$
distribution and fully controlled.

\section{Discussion}

%\indent\indent
Although our study focuses on $h\to W^+W^-$, it raises a broader question of
the potential danger of heavy-flavor leptons in multi-lepton analyses. For
example, trilepton searches for supersymmetry typically have soft additional
leptons.  There could be a significant impact on analyses of these types of
signals if lepton transverse momentum cuts must be raised to remove the
heavy-flavor leptons.

In investigations at linear colliders, the Higgs boson decay $h \rightarrow
W^+ W^-$ can be reconstructed fully from hadronic decays of the $W$ in the
Higgs-strahlung process $e^+ e^- \rightarrow h Z \rightarrow W^+ W^- Z$, with
$Z \rightarrow q \bar{q}$ or $Z \rightarrow l^+l^-$~\cite{Meyer:2004ha}.  The
branching fraction $BR(h \rightarrow W W^*)$ can be measured to $\sim 4$\%
accuracy in $e^+ e^- \rightarrow h Z \rightarrow W W^* Z$, with $W W^*
\rightarrow 4$~jets or $W W^* \rightarrow l \nu + 2$~jets.  Studies of the
decay mode $h \rightarrow l^+l^-E_{\rm miss}$ do not appear necessary for
access to the $h \rightarrow W^+ W^-$ coupling or branching fraction, but
there may be interesting additional information to be gained.  In $e^+e^-
\rightarrow h Z$, with $Z \rightarrow l^+ l^-$, and $h \rightarrow W^+W^-
\rightarrow l^+l^- +E_{\rm miss}$, there should be interesting kinematic
signatures in the four-charged-lepton final state.

\begin{theacknowledgments}

Work in the High Energy Physics Division at Argonne is supported 
by the U.~S.\ Department of Energy, Division of High Energy Physics, 
Contract No.\ W-31-109-ENG-38.

\end{theacknowledgments}


\begin{thebibliography}{9}


%\cite{Sullivan:2006hb}
\bibitem{Sullivan:2006hb}
Zack~Sullivan and Edmond~L.~Berger,
%``Missing heavy flavor backgrounds to Higgs boson production,''
Phys.~Rev.~D {\bf74}, 033008 (2006), arXiv:hep-ph/0606271.
%%CITATION = HEP-PH 0606271;%%

\bibitem{Abazov:2005un}
D0 Collaboration, V.~M.~Abazov, {\it et al.},
Phys.\ Rev.\ Lett.\  {\bf 96}, 011801 (2006).
%%CITATION = HEP-EX 0508054;%%

\bibitem{ATLTDR}
ATLAS Collaboration, ATLAS Technical Design Report Vol.\ II, CERN-LHCC-99-15,
p.\ 704.

\bibitem{PYTHIA}
%\bibitem{Sjostrand:2000wi}
T.~Sjostrand, {\it et al.},  
%P.~Eden, C.~Friberg, L.~Lonnblad, G.~Miu, S.~Mrenna,
%and E.~Norrbin, 
Comput.\ Phys.\ Commun.\ {\bf 135}, 238 (2001);
%[arXiv:hep-ph/0010017].
%\bibitem{Sjostrand:2003wg}
T.~Sjostrand, L.~Lonnblad, S.~Mrenna, and P.~Skands, 
arXiv:hep-ph/0308153.
%%CITATION = HEP-PH 0010017;%%
%%CITATION = HEP-PH 0308153;%%

\bibitem{Carena:2000yx}
Higgs Working Group Collaboration, M.~Carena, {\it et al.}, 
in {\sl Physics at Run II: the Supersymmetry/Higgs Workshop}, Fermilab, 1998,
edited by M.~Carena and J.~Lykken (Fermilab, Batavia, 2002), p.\ 424.
%arXiv:hep-ph/0010338.
%%CITATION = HEP-PH 0010338;%%

\bibitem{ATLLEPS}
ATLAS Collaboration, ATLAS Technical Design Report Vol.\ I, CERN-LHCC-99-14;
with updates from B.\ Mellado, S.\ Paganis, W.\ Quayle, and
Sau Lan Wu, ATL-CAL-2004-002, unpublished;
F.\ Derue, C.\ Serfon, ATLAS-PHYS-PUB-2005-016, unpublished.
% Muons checked vs. Higgs working group slides, 12/14/2005.

\bibitem{Maltoni:2002qb}
F.~Maltoni and T.~Stelzer, JHEP {\bf 0302}, 027 (2003).
%%CITATION = HEP-PH 0208156;%%

\bibitem{Meyer:2004ha}
N.~Meyer and K.~Desch,
Eur.\ Phys.\ J.\ C {\bf 35}, 171 (2004);
%\bibitem{Garcia-Abia:2005pf}
P.~Garcia-Abia, W.~Lohmann, and A.~Raspereza,
Eur.\ Phys.\ J.\ C {\bf 44}, 481 (2005); 
%\bibitem{Aguilar-Saavedra:2001rg}
ECFA/DESY LC Physics Working Group, J.~A.~Aguilar-Saavedra, {\it et al.}, 
arXiv:hep-ph/0106315.  
%%CITATION = EPHJA,C35,171;%%
%%CITATION = HEP-PH 0106315;%%
%%CITATION = EPHJA,C44,481;%%

\end{thebibliography}
\end{document}